# Who Gets What, According to Whom? An Analysis of Fairness Perceptions in Service Allocation

Jacqueline Hannan
jmhannan@buffalo.edu
University at Buffalo
Buffalo, NY, USA

Huei-Yen Winnie Chen
winchen@buffalo.edu
University at Buffalo
Buffalo, NY, USA

Kenneth Joseph
kjoseph@buffalo.edu
University at Buffalo
Buffalo, NY, USA

## ABSTRACT

Algorithmic fairness research has traditionally been linked to the disciplines of philosophy, ethics, and economics, where notions of fairness are prescriptive and seek objectivity. Increasingly, however, scholars are turning to the study of what different people *perceive* to be fair, and how these perceptions can or should help to shape the design of machine learning, particularly in the policy realm. The present work experimentally explores five novel research questions at the intersection of the "Who," "What," and "How" of fairness perceptions. Specifically, we present the results of a multi-factor conjoint analysis study that quantifies the effects of the specific context in which a question is asked, the framing of the given question, and who is answering it. Our results broadly suggest that the "Who" and "What," at least, matter in ways that are 1) not easily explained by any one theoretical perspective, 2) have critical implications for how perceptions of fairness should be measured and/or integrated into algorithmic decision-making systems.

Scholarship within the algorithmic fairness community has traditionally focused on identifying and formalizing mathematical definitions of fairness. This work has developed a robust suite of quantitative and at least seemingly objective definitions of fairness that match perspectives in law, philosophy, statistics, and ethics [18]. However, rather than seek a single mathematical definition of algorithmic fairness, scholars have more recently begun to understand it as a subjective quantity [9, 12, 26, 35, 41, 43, 44]. This psychological perspective, as psychologists Van den Bos and Lind put it, defines fairness as "an idea that exists in the minds of individuals… This subjective sense of what is fair or unfair…can be contrasted with objective principles of fairness and justice that are studied by philosophers, among others" [39, pg.7].

From this perspective, then, the goal of algorithmic fairness research shifts from defining the right "objective" measure of fairness to understanding, acknowledging, and addressing the myriad perspectives that exist within the sociotechnical system into which an algorithm will be placed [12, 44]. For example, scholars have looked at how perceptions of fairness in automated decisions vary across application contexts [35], who is being asked [41], and approaches to explaining what the algorithm is doing [5]. In turning towards these questions, the study of algorithmic fairness connects with a large literature in the social sciences. In political science, for example, the widely acknowledged *deservingness heuristic* states that people are generally motivated to believe it is fair to allocate public goods to "deserving" others, i.e., people who need the service because they are unlucky and not because they are lazy [40].

There is, however, an important distinction between work in the social sciences and in the AIES/FAccT community. Namely, the AIES/FAccT community is concerned not only with what people believe to be fair or unfair, but also how these beliefs can inform the development of automated decision-making (or decision-assistive) technology. That is, many AIES/FAccT scholars are interested in how we can take measurements on what a population considers to be (un)fair and incorporate them into, e.g., the automated allocation of a public good [10, 22], risk assessments in the criminal justice system [21], or the decision of a self-driving car [4, 27]. While such approaches hold the potential to equalize the ways in which AI distributes power [22], these efforts make it all the more critical to understand the sources of variation in our measurements of fairness perceptions, and how these sources of variation may lead to different measurements in different contexts and across different populations.

Our work is driven by five research questions that address these sources of variation in measurements of fairness perceptions. As in prior work, we operationalize fairness perceptions as the ways in which survey respondents choose to allocate (or remove) services (or resources) from different hypothetical individuals. While such allocation tasks do not cover all contexts, they are perhaps the most common setting in which fairness has been studied. Specifically, we address questions pertaining to how fairness perceptions vary based on 1) the particular service being allocated or 2) demographics of the respondent, and 3) how questions that theoretically target different cognitive mechanisms can impact measurements.

To address these questions, we develop and carry out an extensive *conjoint analysis* [14] survey experiment on a large, U.S.-centric sample of Mechanical Turk workers. Examples of the main types of questions we ask survey respondents are presented in Figure 1. Conjoint analysis is widely used across other disciplines, in particular political science and healthcare, to identify the attributes people use to make decisions. Indeed, others have used it in similar analyses of fairness perceptions [4, 32]. This wide use is because conjoint analysis has been shown to reliably identify factors relevant to real-life decision-making in ways that more reflective ways of asking survey questions, e.g. Likert-scale responses, do not [14, 23]. For example, Hainmueller et al. [13] validated conjoint analysis and vignette-based surveys against real-world behavior. Their findings show that the conjoint method comes within 2 percentage points

You are a social worker that provides a small tuition stipend to individuals who want to take courses at the local community college.
If you had to choose, which of the two individuals would you say it is most fair to provide the stipend to?

| Attribute | Person A | Person B |
|---|---|---|
| Age | 20 | 70 |
| Children | Two kids | Two kids |
| Occupation | doctor | nurse |
| Criminal History | Prior history of non-violent crime | Prior history of violent crime |
| Race | Hispanic | Hispanic |
| Upbringing | Grew up poor | Grew up middle class |
| Political Affiliation | Democrat | None specified |
| Health Issues | Generally Healthy | Mental health issues |

[ Person A ]  [ Person B ]

You are a doctor at the height of the COVID pandemic. Two patients are currently on ventilators, but you need one of the ventilators for an ill child. Without a ventilator, the patients may die.
If you had to choose, which of the two individuals above would you **remove** the ventilator from?

| Attribute | Person A | Person B |
|---|---|---|
| Upbringing | Grew up upper class | Grew up middle class |
| Race | Black | Black |
| Health Issues | Generally Healthy | Generally Healthy |
| Political Affiliation | None specified | Republican |
| Age | 40 | 40 |
| Criminal History | Prior history of non-violent crime | None |
| Children | No kids | No kids |
| Occupation | janitor | nurse |

[ Person A ]  [ Person B ]

Figure 1: Two examples (top and bottom) of questions asked to respondents of our survey experiment. The top is from the Low Severity, Social work service setting, and the Reward x Fairness Cued question framing condition. The bottom is from the High Severity, COVID setting, and the Punish x non-fairness cued question framing condition. See below for details.

of the behavioral benchmarks used in their study, implying that a conjoint design replicates a true decision-making situation more accurately than a vignette or direct survey-based approach.

Our analysis results in 4 main findings:

- *What* **resource is being allocated is important** - Perceptions of who should get services varied in both predictable and unpredictable ways. For example, predictably, it was perceived to be fair to consider a person's upbringing when allocating affordable housing, but not a COVID medication. Unexpectedly, however, respondents were *more* likely to give old people (relative to young people) palliative care, but *less* likely to give them to be given life-saving resources.
- *Who* **allocates the resource is important** - Survey respondents' fairness perceptions varied along partisan lines. These variations manifested in ways that emphasize group-justifying preferences (e.g. Democrats give more to Democrats) and ideological differences (e.g. Democrats are more inclined to consider an individual's upbringing).
- *How* **one asks about fairness perceptions is less important** - Different ways of asking (what seem like) the same question can produce different results, but rarely. In particular, question framing seemed to have effects only in low-stakes allocation settings, and questions that triggered reflexive cognition were broadly consistent with those targeting more subconscious mechanisms.
- **Racial inequality consistently manifests in our analyses, regardless of What, Who, or How**- While respondents almost universally treat race, when explicitly stated, as a variable to be ignored, racial biases existed in fairness perceptions via indirect paths related to systemic racial inequality in the United States, specifically in the context of the criminal justice system. These effects were observed across all services, question framings, and respondent political affiliations.

Taken together, our findings imply that perceptions of fairness are difficult to know *a priori*, and thus empirical measurement in new settings of "What", "Who," or "How" are critical. At the same time, however, our results can be explained by a mix of existing social theory, suggesting continuing efforts to marry social theory, subjective fairness, and automated decision making are critical [12, 22]. To encourage further work along these lines, all code and data used in this study are available at https://github.com/kennyjoseph/fairness_perceptions_aies.

## 1 BACKGROUND AND MOTIVATION

A host of theories exist to explain how individuals make decisions about what is fair and/or who deserves certain (usually public) goods or services. Here, we present four representative perspectives that help to shape our study design and analysis, and where they have been leveraged in prior work on algorithmic fairness. Note that absent from our analysis here are discussions about philosophical– or what Grgic-Hlaca et al. [12] call "prescriptive"– notions of fairness, where one seeks to identify the "right" definition. Rather, we here focus on what is known about *why people are likely to have certain views on what is fair*, linking the discussion of our results to the prescriptive models.

Perhaps the most straightforward view is that people are more likely to allocate to others that hold similar, salient social identities. This perspective– that Jost [19] calls the *ego-justifying* or *group-justifying* perspective– aligns with both notions of self-interest and in-group favoritism. However, sometimes, neither self-interest nor in-group favoritism can readily explain preferences for the allocation of public goods. For example, Americans tend to support capitalist notions of the unequal distribution of wealth, even when they know it will not benefit them or their group. In response to these observations, Jost and Banaji [20] developed *Systems Justifications Theory* [for a recent review, see 19]. Systems Justification Theory proposes the idea that individuals, often subconsciously, are motivated to respond to the distribution of goods and services in ways that reflect their perceptions of the status quo. This tendency towards the status quo leads us to allocate goods and services in a way that maintains existing social structures and hierarchies.

In Political Science, the *deservingness heuristic* presents a similar perspective to systems justification [31]. The deservingness heuristic focuses on the (theorized) evolutionary habit of humans to prefer allocating services to those who are unlucky, rather than those who are lazy. Thus, the theory assumes that when allocating services, we first categorize others based on whether (we believe) they are unlucky or lazy. We then use this categorization to determine whether a person is deserving or not. Lazy people are undeserving, because they have put in no effort that should be rewarded, whereas unlucky individuals are deserving because the circumstances resulting in a need for resources are perceived to be out of their control [30]. Importantly, while the deservingness heuristic suggests that resource allocation should be consistent across public services given knowledge of how people categorize others as unlucky or lazy, it leaves room for the fact that these categorization processes may differ substantially across a population [1].

Finally, economists have developed a similar model around the idea of Equality of Opportunity (EOP) [33]. The EOP model states that individuals should be treated similarly, regardless of artificial social barriers. The model breaks an individual's attributes down into 2 categories: *circumstance* — those that should not impact judgement about the individual (e.g. race) – and *effort* (those that should, e.g. need). EOP models assume that the expected utility for individuals with similar effort should not be impacted by changes in circumstance. Yaghini et al. [44], adopt EOP to study fairness perceptions in work closely related to ours; we return to this work in our Discussion section.

Collectively, these four perspectives of how individuals make decisions about the allocation of goods and services present two critical ways in which fairness perceptions may vary. First, both Systems Justification Theory and the EOP model suggest that *the service being allocated should impact fairness perceptions*. Systems Justification Theory makes what is essentially an assumption about *prototypicality* - that is, we should expect that an individual will allocate services to those who currently (are perceived to) get the service. In contrast, EOP allows for the assumption that the features utilized across service allocation conditions may be consistent (the "circumstance" is a consistent set), but that the weight of the "effort" variables can change across contexts. Thus, these theories make different mathematical but similar conceptual arguments about the importance of individual features in perceptions of how a service should be allocated.

However, we are unaware of any studies that consider these theoretical predictions by testing them across different service allocation settings within a single experiment. The closest work we are aware of is that of Srivastava et al. [35], who study different perceptions of fairness across different contexts, but focus on different mathematical definitions of fairness as opposed to studying the attributes respondents use in their decisions. This leads to our first research question:

**RQ1:** *How do fairness perceptions vary across different services that are allocated?*

Second, the ego- and group-justifying perspective and the deservingness heuristic imply that there should be variation across *respondents* in the ways that services are allocated. This has been observed in other related work. For example, Yaghini et al. [44], using an EOP-style analysis, find that respondents with different ages and genders vary in their view of what features should be considered circumstance versus effort-related. And Awad et al. [4] observe country-level variation in perceptions of life-and-death settings for self-driving cars. However, it remains to be seen how such differences emerge across a range of different service decisions. This leads to our second research question:

**RQ2:** *(How) do fairness perceptions vary across respondents with different demographics in the context of service allocation?*

In addition to these two theoretical dimensions of variation, there are also important methodological factors that may change the way that individuals respond. One important question is whether or not individuals who are primed to think about fairness specifically, rather than just who "should" get a given service, respond differently to questions about service allocation. AIES/FAccT scholars have used question wordings that sometimes cue fairness and sometimes do not. For example, Saxena et al. [34] asks participants to select the individual who should receive a loan payment from a pair, whereas Harrison et al. [15] instructs participants to rate the fairness level for a particular individual receiving bail.

However, there is reason to believe that prompting individuals to think about fairness (or not) might impact their decision-making, by impacting the extent to which respondents use "fast" thinking, driven primarily by emotional, reactive, gut responses, and "slow" thinking, driven by more cognitive, reflective responses, depending on the situation [38, 39]. In the latter case, respondents may reflect on, e.g., institutional norms, driving their responses to reflect what is socially desirable [25]. Given the normative associations with the concept of fairness, we might expect that individuals use a more "gut reaction" when asked who *should* get a service, but are primed to be more reflective (and thus change their feature preferences) when asked to allocate *fairly*. This leads to the following research question:

**RQ3:** *How do fairness perceptions vary when measured using questions that do or do not cue people to think about fairness?*

Similarly, algorithmic fairness scholars have looked at both cases where judgements result in punishments, e.g. in the context of recidivism [11, 12, 26], and where services are rewards, e.g. in hiring settings [24]. Once again, there is reason to suspect that this subtle difference may result in different ways that individuals respond. Specifically, in punishment contexts, respondents are primed with a negative connotation, and may therefore be prompted to look for "issues," whereas services-as-benefits may lead individuals to look towards aspects of people that emphasize positive qualities. This potential discrepancy leads to the following research question:

**RQ4:** *How do fairness perceptions vary when measured using questions that prompt punishment (e.g. service removal) versus reward (e.g. service provision)?*

Finally, existing work also varies widely on how responses are solicited. Scholars in the AIES/FAccT community have primarily leveraged three different approaches to soliciting perceptions of fairness. First, several scholars have taken a feature-specific approach, asking on a Likert scale how important a given feature was in their decision-making. For example, Yaghini et al. [44] asked participants which parameters are acceptable to consider when making a recidivism prediction, such as race or criminal record. And Grgic-Hlaca et al. [12] ask respondents survey questions about both which features are fair, and properties of those features. Second, scholars have relied on what we will call forced choice hypothetical individual decisions, where participants are asked whether or not they would allocate a particular good or service to a single individual or to choose between several individuals. Perhaps the most well-known example of this is the work of Awad et al. [4], whose *Moral Machine* experiment employed this design to study moral decisions in a trolley problem-like setting with self-driving cars, but other examples exist [21]. Finally, several works have focused specifically on fairness measure decisions, where participants are asked to choose an established, mathematical definition of fairness that they prefer [15, 35].

These different methods have different mathematical aims, but also, again, trigger potentially distinct cognitive mechanisms. For example, forced-choice designs used by AIES/FAccT scholars derive from approaches that specifically aim to trigger subconscious preferences that align with behavior [14], whereas Likert-scale responses are known to encourage more reflection on the specific feature or decision criterion. This difference is related to the distinction between what people think they believe, and what they actually do. As an example, consider that individuals on dating websites are likely to state that they have no racial preferences on questionnaires, but empirically show strong racial preferences. When asked directly to reflect on race, individuals largely invoke social norms that dictate one *should* not have racial preferences. But when selecting partners on the site, individuals are largely driven by their faster, more reactive thinking, where stereotypes and prejudices are more likely to be reflected [3].

As such, we expect there may be differences in the ways in the importance respondents place on different features based on how we inquire about that feature, specifically, in the difference between force choice responses from questions like those in Figure 1, and the feature-specific Likert-scale questions in prior work. This leads to our final research question:

**RQ5:** *How do different features vary in their impact of fairness judgements depending on how the response is solicited?*

## 2 EXPERIMENTAL DESIGN

Our study was conducted with 747 participants on Amazon Mechanical Turk (MTurk), see below for further details on the participant sample. The study was approved by the Institutional Review Board at the University at Buffalo.

After consenting to the study, participants first saw a tutorial introducing them to the types of questions asked in the survey. The tutorial included three comprehension questions. If the participant could not correctly answer the three questions, they were not allowed to continue. The main portion of the survey is an experiment that asks ten questions, like those in Figure 1. We vary both the question prompt (i.e. what service the respondent is cued with and how the question is worded) and whom the respondent is asked to decide between (i.e. who Person A and Person B are in Figure 1).

With respect to the question prompt, we construct a 4-way, 2x2x2x2 fully factorial experiment. The first two factors are relevant to the *service setting* (to address *RQ1*), and the second two to the phrasing of the question (to address *RQ3* and *RQ4*). Participants are placed randomly into one of 16 conditions of the full-factorial design, meaning that a study participant is asked about only one type of service and shown one question phrasing for all ten survey questions. Additional details on the factorial design for the question prompt are given below.

We then explore the preferences that respondents have on whom to allocate services to, and how these are impacted by the prompt given. We do so by first identifying a set of possible *attributes* (e.g. Age) and then a set of *attribute levels* (e.g. 20, 40, and 70 years old). Table 1 provides the attributes and attribute levels that we consider in our study. We then construct two hypothetical individuals, who receive a random level for each attribute. In Figure 1, these hypothetical individuals are "Person A" and "Person B." Survey respondents are then asked to choose whom between the two individuals would be given the service or have it taken from them (depending on the prompt).[1] This random sampling approach allows us to *causally* interpret the impact of different attribute levels on respondents' service allocation decisions. Full details on this part of the approach to identifying respondent preferences across attributes are given below.

After the survey experiment, respondents answered a set of Likert questions asking them how much they believed they weighted each attribute in their decision (to address *RQ5*). Specifically, for each attribute, we asked respondents to "indicate how important [this attribute was] to you in your earlier decisions about allocating services." Finally, respondents answered a standard battery of demographic questions for age, race, gender, political affiliation, education, and prior experiences with the services being allocated. In the present work, we focus only on respondent's political affiliation, however, all demographics are provided in the data release for this study.

### 2.1 Factorial Experiment Design

We refer to the question prompt shown to respondents as their *decision context*. When designing the parameters that defined the decision context, two main categories are taken into consideration. First, to address *RQ1*, we vary the *service setting*, i.e. we consider which service is being allocated. To do so, we vary two factors- a service context (relating to the high-level institutional context in which the service is allocated), and a service severity level (relating to the importance of the resource available). Second, we consider variations on *question framing* to address *RQ3* and *RQ4*. Full prompts for all 16 conditions are given in the code release for this paper; Figure 1 provides examples of two conditions, and the caption notes which conditions the two sample questions are drawn from.

---

[1]Respondents were, in some cases, also provided a response on a slider to indicate how confident they were in their response. Analyses of respondent confidence are not considered in the present work, however, data on the sliders are provided in the data release for this paper.

| Attribute | N | Attribute Levels |
|---|---|---|
| Race | 4 | White, Black, Asian, Hispanic |
| Age | 3 | 20, 40, 70 |
| Criminal History | 3 | None, Prior history of non-violent crime, Prior history of violent crime |
| Health Issues | 4 | Generally healthy, Mental Health Issues, Physically Disabled, Diabetic |
| Occupation | 15 | janitor, nurse, doctor, unemployed, firefighter, Instacart shopper, artist, politician, banker, scientist |
| Children | 2 | No kids, 2 kids |
| Upbringing | 3 | Grew up poor, Grew up middle class, Grew up upper class |
| Political affiliation | 3 | Democrat, Republican, None Specified |
| | | 25,920 possible combinations |

Table 1: The eight attributes allotted to hypothetical individuals in our conjoint analysis design.

The two service contexts used are social work and COVID-19. Within each service context, two possible situations are created, one of high severity and one of low severity. In the case of social work, the high severity situation involves *housing availability* for people without a place to live, and the low severity situation relates to *tuition assistance* for courses at a local community college. For COVID-19, the high severity scenario involves a *life-saving device*, i.e. a new treatment that improved survival odds in a COVID patient by 50% or the use of a ventilator, and the low severity scenario involves a *pain reduction* medical device that alleviates mild discomfort that some COVID patients experience.

With respect to question framing, to address *RQ3*, the question is framed as either a gain or loss of a service and written in order to either cue thinking about punishment or benefit. In the gain scenarios, the resource for the given context is *given* to either person A or person B. In the loss scenarios, the respondent is asked to *take* the resource from either Person A or Person B.[2] To address *RQ4*, we either do or do not cue respondents to think explicitly about fairness. In the case of cuing fairness, the question is phrased to ask which outcome is "most fair." In the absence of a fairness cue, the respondent is simply asked which individual they would *give* the specific service to.

## 2.2 Identifying Respondent Preferences

While we are interested in variation along all combinations of attributes, it would be infeasible to ask a sufficient number of respondents about pairwise comparisons across all 25,920 possible combinations of the attributes we study. Instead of attempting a full factorial design or removing conditions, our experiment therefore uses a random sampling approach and applies principles from conjoint analysis [14] to derive effects of the different factors. Essentially, we show survey participants hypothetical individuals that are randomly generated via sampling in a principled, well established experimental approach.

To generate each question for each respondent in a conjoint analysis, we perform a two step procedure. First, we generate a random ordering of the attributes, to protect against ordering effects. Second, we randomly draw an attribute level for each of the two hypothetical individuals.[3] Respondents are then asked to select between these respondents based on the question prompt given.

The attributes used for this study are chosen to reflect our research questions and the expected outcomes of our study, based on previously mentioned theories in the social sciences. Preferences on the attributes selected can play a role in reflecting these theories. For example, the notion of political affiliation could help identify any possible group-justifying behavior. And we anticipated other variables, in particular upbringing, criminal history, and health, to display effects aligning with the deservingness heuristic.

## 2.3 Participant Population

As noted, we use MTurk to recruit participants for our study. Because service allocation perceptions are already well-known to vary widely across national culture [1, 4], we restrict the present analysis to only respondents in the United States. In order to ensure this was the case, we leverage a tool provided by [42] to ensure that participants' IP addresses are located within the United States and that they are not using a VPN.

While there are acknowledged issues with Mechanical Turk respondent pools, a range of work suggests that it is a practical sample for the exploratory psychological analyses conducted here [7, 29] Indeed, overall, studies have found data produced using MTurk to be comparable and as reliable as data obtained using the typical American college samples and Internet users [7, 28]. However, scholars have also noted ways to help ensure quality responses from MTurk. We use three of these approaches in the current study. First, we sought to ensure workers are reasonably well-paid; based on estimates of completion time, we expect workers are paid around $12.50 per hour. Second, we request responses only from Turkers who had greater than a 95% completion rate and had completed over 1,000 HITs. Third, we use comprehension questions to ensure task competence and, more importantly, to remove bot-like responses.

In total, the survey was started by 1,018 respondents, but only 747 (73.3%) passed both the IP filter and our task comprehension checks. The median age of respondents is 35. Our respondents lean left politically; 58% identified as a Democrat or leaning left, compared to 31% Republican/lean right, and 10% Independents/no leaning (1% do not provide political affiliation). Of respondents who provided education, 12% have a postgraduate degree, 55% have a two or four year college degree, and 33% have a high school diploma or less.

---

[2]Note that the resource is consistent in the gain vs. loss situations for social work, but varies slightly in the COVID-19 high severity context, for the sake of logic. For the gain condition, the allocated resource is a treatment that improves survival odds by 50%. For the loss condition, the resource that must be removed from one individual is a ventilator.

[3]These levels can potentially match, but we resample under the (unlikely) condition where both hypothetical individuals are exactly alike.

Finally, our sample skews White, 76% of respondents said they are White, compared to only, e.g. 7% Latinx and 11% Black.[4]

In sum, our sample is not representative of any clear population, and thus, as we emphasize below, one should be cautious about generalizing specific results on specific attributes to any population beyond the study participants. In contrast, we focus mainly on more generalizable effects of our experimental manipulations and demographic differences that align with existing theory.

## 3 RESULTS

### 3.1 *RQ1*: Variation across Service

We find that the service setting respondents are given had a significant impact on their responses, but that these impacts 1) are not systematic across the different attributes and 2) often are a-priori non-obvious for particular attribute levels. Figure 2 presents marginal mean estimates for each level of each attribute, with the different service settings presented in varying colors and point shapes. Notably, effects do not consistently vary across the main effects of our 2x2 service context by severity experimental design, and thus we consider each cell of the 2x2 separately. Marginal means in conjoint analysis experiments are a causal quantity that represent the probability of a respondent giving (or not taking away) a service to (or from) an individual with the certain attribute level, relative to any other level, for a given service condition [23]. For example, the purple cross in the "Age" subplot can be read as, "Respondents gave (took away from) 70 year olds significantly less (more) tuition assistance than would be expected by chance, allocating the assistance to 70 year olds only around 40% of the time." Note that despite the different number of levels across factors, marginal means are always estimated relative to a base rate of being selected in a pairwise comparison, and thus the base (chance) rate is always 50%.

Across all service settings, respondents most consistently rely on age, criminal history, and whether or not an individual has kids. The direction of effects for having kids and having a prior criminal history are consistent across contexts; respondents were always more likely to give to individuals with kids, no criminal history, and a history of only non-violent crime relative to a history of violent crime. In contrast, while age is a consistently strong predictor, the direction of effects vary across service setting. Older individuals are significantly more likely to be given pain reduction medicine or housing, and significantly less likely to be given tuition assistance or a life-saving device.

Effects for attribute levels of the other variables are less consistent. Political affiliation is generally a weak predictor across contexts; however, as we discuss next, this effect is to some degree masked because of group-justifying behavior. With respect to health issues, respondents are much more likely to provide housing to the physically disabled, less likely to provide life saving devices to those with mental health issues but more likely to provide the device to those with diabetes, and less likely to give healthy people housing or pain medicine. Respondents largely do not factor explicit mentions of race into their decisions, although Black people are slightly more likely to be given tuition assistance. Finally, significant preferences in allocation across occupations are sporadic.

---
[4]We did not request information about the respondent's gender because we had no reason to believe this was of theoretical interest to the present work.

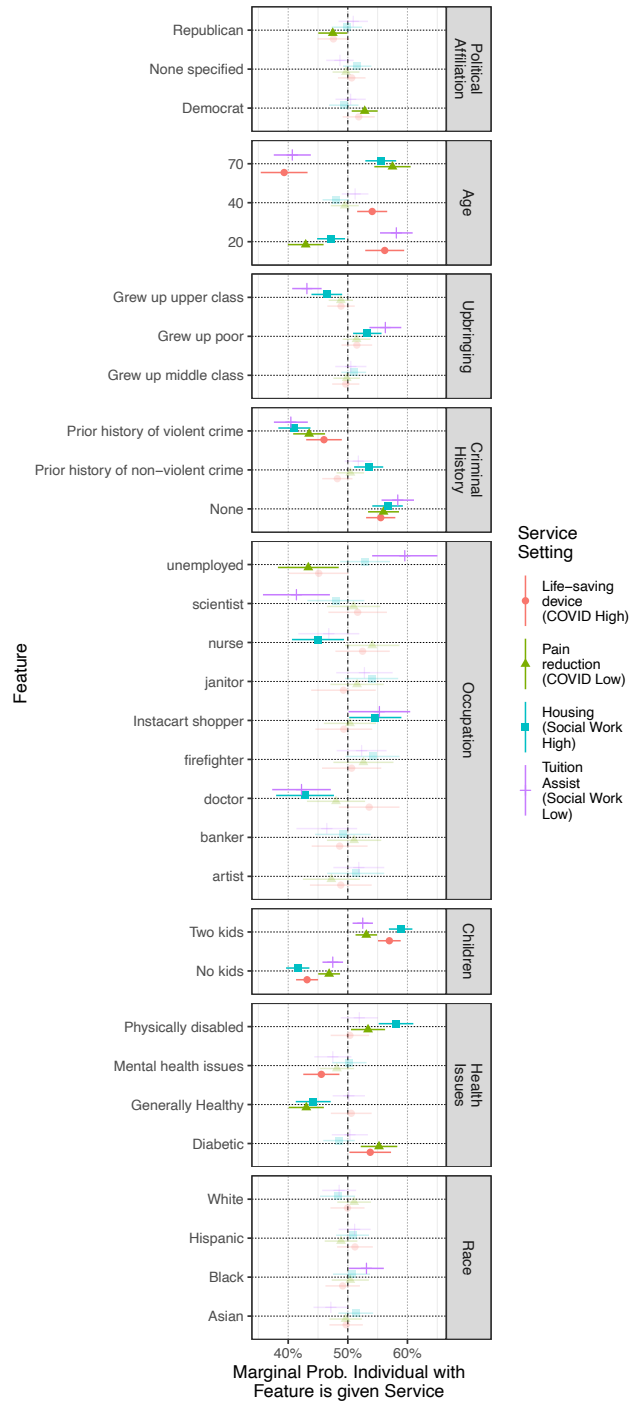

Figure 2: The marginal probability (x-axis) that a hypothetical individual having a given attribute value (y-axis) is given a particular service. Services are differentiated by color and shape. Values not significantly different from chance ($\alpha = .05$) are dimmed. Confidence intervals are 95% standard errors, controlling for multiple observations of respondents. Different attribute types are separated into different subplots.

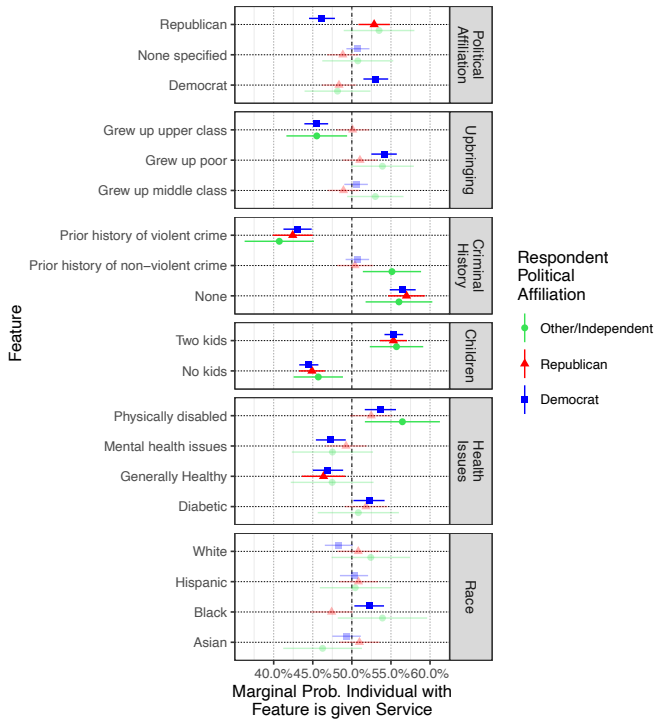

Figure 3: The marginal probability (x-axis) that a hypothetical individual having a given attribute value (y-axis) is given any service, based on the respondent's political affiliation (different color/shaped point ranges).

Perhaps more telling for occupation is where results are unexpectedly null. In particular, respondents are no more likely to give doctors or nurses assistance in the COVID settings than individuals with other occupations, although doctors are significantly more likely to get COVID assistance than social work services.

In sum, Figure 2 tells a complex story at the attribute level. However, three overarching findings are worth summarizing. First, the service being allocated clearly matters. For example, our results show that one should not expect that the ways in which particular attributes impact fairness perceptions in life saving or life defining situations remain the same in less critical situations. Second, and in contrast, is that service setting does not always matter. In particular, the importance of criminal history is relatively consistent across the four different scenarios studied. Finally, specific preferences at the attribute level are difficult to predict a priori. For example, it is not immediately obvious why respondents disliked the idea of older people suffering from pain or homelessness, but felt they should not live or be given an opportunity for education at the expense of a younger individual.

## 3.2 *RQ2*: Variation across Respondent

We find significant evidence of group-justifying effects on service allocation, in that respondents prefer to allocate services to individuals who share their partisanship. Figure 3 shows the same information as Figure 2, except that results are now aggregated across respondent's political affiliation instead of service setting.[5]

We find that Democrats allocate almost 10% more services to other Democrats, and Republicans allocate around 5% more to other Republicans. For Democrats this difference is driven not only by in-group favoritism – a preference for allocation to Democrats – but also by "out-group hate" [6]– a term coined to denote distaste for allocating services to Republicans .

However, differences between individuals aligning with these political parties do not stop at a preference (distaste) for (dis)similar others. Most notably, Democrats are significantly more likely to consider class, or upbringing, than Republicans. Democrats are almost 10% more likely to allocate services to individuals who grew up poor, relative to those who grew up upper class, whereas no significant difference existed for Republican respondents. Democrats are also slightly more likely to allocate services to Black people; Republicans are not. These results show that differences between respondents move beyond a preference for those with the same identities to ideological differences as well.

While these differences are important, it is also worth noting significant similarities also exist across parties. In particular, respondents across the political spectrum have relatively consistent views on criminal history and having (or not having) children. The primary difference across the spectrum on these variables is from Independents, who showed no preference for individuals who committed no crimes relative to those who committed only non-violent crimes. Racial perspectives, too, were similar across the political spectrum outside of the difference in views towards Black people.

## 3.3 *RQ3-4*: Variations on Question Structure

We find inconsistent evidence that manipulations to whether or not respondents are cued with fairness (*RQ3*) or primed to think about punishment versus reward (*RQ4*) had strong effects on respondent preferences, especially in high-severity settings. Table 2 shows results from F-tests to evaluate whether or not differentiating responses by question framing manipulation significantly improves the explainability of regression models of conjoint responses. These F-tests represent the standard approach to evaluating the importance of potential moderating variables in conjoint analysis [23].

| Framing | Service Condition | Fval | p |
|---|---|---|---|
| Fairness Cue (*RQ3*) | Pain reduction | 1.45 | 0.07 |
| | Life-saving | 0.99 | 0.48 |
| | **Tuition Assist** | **1.62** | **0.03** |
| | Housing | 0.74 | 0.82 |
| Punishment vs. Reward (*RQ4*) | **Pain reduction** | **1.64** | **0.03** |
| | Life-saving | 1.37 | 0.11 |
| | **Tuition Assist** | **2.20** | **< .001** |
| | Housing | 1.31 | 0.14 |

Table 2: Results of F-test on significance of changing framing from Gain to Loss, or from "should get" to "fair" (i.e. for RQ3 and RQ4). The degrees of freedom for all tests is 24. Rows are bolded if p-values are significant at $\alpha = .05$

---

[5]We do not show results for age, due to the contrasting impacts across context, and also do not show occupation for space reasons. These results can be found in the code and data release for this article.

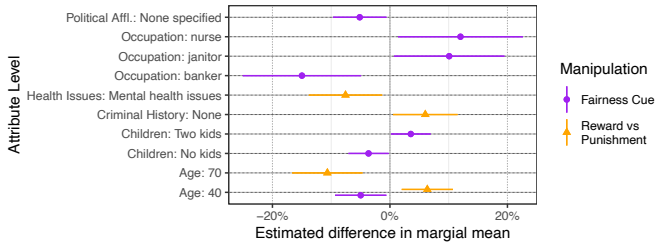

**Figure 4: Effects of question manipulations for *RQ3* (purple) and *RQ4* (orange) for attribute levels where the manipulation results in a significant difference across conditions ($\alpha = .05$). For *RQ3*, results represent mean + 95% CI (standard error) estimates of the increase in odds than an individual gets a service if fairness is cued. For *RQ4*, results represent an increase or decrease if reward, rather than punishment, is cued.**

We find that *RQ3* and *RQ4* manipulations only have a significant effect ($\alpha = .05$) for 3 out of the 8 possible context/manipulator combinations, and in only one of those conditions is $p < .01$.

We do find, however, that both manipulations have significant overall effects for the condition where tuition is allocated (low severity, Social Work context), and the Punishment vs. Reward has an effect in the Pain Reduction setting (low severity, COVID context). Our results therefore leave open the possibility that decisions are more sensitive to the way things are asked in less critical contexts.

Even here, however, effects on specific attribute levels are not obviously systematic. Figure 4 presents attribute levels for the two research questions where the manipulation has a significant effect on responses. With respect to the fairness manipulation, respondents are more likely to provide individuals with children with tuition, as well as nurses and janitors, and less likely to give individuals with no political affiliation, bankers, or middle-aged individuals, when cued with fairness. When cued with reward (versus punishment), respondents are less likely to provide tuition to those with mental health issues or older individuals, and more likely to reward individuals with no criminal history.

### 3.4 *RQ5:* Variations on Question Type

We find evidence that respondents are generally good at inferring which attributes they relied on in the conjoint analysis, although this was not the case for Health Issues. Figure 5 plots two measures of importance for the different attributes in our study, with results split on the political affiliation of respondents. Results in purple represent the importance respondents stated for that variable using the Likert questions given to them after the conjoint experiment. Results in orange represent the importance of each attribute to respondents in the conjoint analysis, as estimated by the average amount of variation in responses explained by the variable. As is standard, we use McFadden's Pseudo-$R^2$ as a measure of variation explained, implemented in the dominanceAnalysis [8] package in R [37]. Note that this measure accounts for *overall* variation explained in ways that may not reflect significance levels for specific attribute levels in Figure 2. For example, we see that Occupation explains more variation in the data than Political Affiliation, even

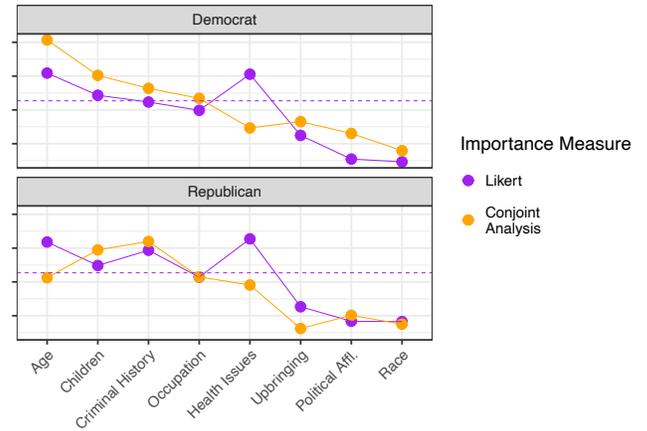

**Figure 5: The y-axis represents estimates for the importance of attributes to respondents. For Likert-scale questions (purple), this is estimated by the respondents themselves (average Likert response). For conjoint responses (orange), this is estimated by the average importance of variables in regressions predicting responses. Results are split by respondent political affiliation, and the two measures of importance are standardized so they can be plotted together. The horizontal purple dashed line represents a response of "Neutral" on the 7-point Likert scale, and is given for reference. Results are computed separately for each of the four service contexts, and then averaged together.**

though we see some significant effects for specific levels of Political Affiliation in Figure 2.

Both Democrats and Republicans respondents were consistent in their perception that Upbringing, Political Affiliation, and Race are not important to their fairness decisions in both their responses on the conjoint analyses and their reflections in the Likert responses. However, we do not observe this agreement for all variables. Specifically, respondents consistently state that they believe they rely more heavily on Health Issues than we find in their behavior in the conjoint analyses. Although not shown here, we find this is driven heavily by the life-saving device service setting, where respondents state a strong reliance on Health Issues, but did not heavily rely on the attribute as a whole in their conjoint responses.

Overall, Democrats are better than Republicans at identifying which variables they relied on in the conjoint analysis. This claim is based on an analysis of the correlation between Democrats' and Republicans' stated view of how heavily they rely on attributes and their actual use as measured in the conjoint analysis. Using Spearman's rank correlation metric, we find a large and significant relationship between these two quantities for Democrats (.76, p=.037). In contrast, there is no significant correlation (.55, p=.17) between these two quantities for Republicans.

Finally, while Democrats and Republicans agreed on what *not* to use, there is less consistency in the relative importance of what was used. Setting aside the outlier of Health Issues, the most important attributes for Democrats according to both measures are, in order,

Age, Children, Criminal History, and Occupation. Republicans, relied on these same factors, but ordering varied from Democrats and across importance metrics. Republicans perceive themselves to rely, in order, on Age, Criminal History, Children, and then Occupation. We find, however, in the conjoint analysis, that they rely, in order, on Criminal History, Children, Age, and then Occupation.

## 4 DISCUSSION

Our results do not align neatly with any one theoretical understanding of how individuals perceive fairness. For example, the ideological differences we find for *RQ2* fit with the deservingness heuristic, which presumes that ideology drives categorization processes that lead to service allocation decisions based on whether one is perceived to be lazy or unlucky. But other results do not line up along the lazy versus unlucky divide. In particular, it is unclear from a deservingness perspective why either having children or being old would be a significant factor, despite these two representing the most important preferences driving respondents' fairness perceptions in our study. These results, instead, perhaps align better with a Systems Justification perspective, which itself explicitly discounts the observed group-justifying observations in Figure 3.

Rather, for any given respondent and service setting, our results suggest we should expect to require a mix of theoretical perspectives to understand fairness perceptions. This, in turn, points to a need for theoretically-driven empirical approaches to identifying fairness perceptions across the rich arena in which algorithms are now making policy decisions. Several important examples already exist for how to do such work. In sociology, Hoey and MacKinnon [17] presents a framework for understanding social decision-making through a mix of empirical estimates derived from surveys and theoretically informed parameter settings. More in line with the algorithmic fairness work, Grgic-Hlaca et al. [12] presents a model in which they assume that features are drawn from a distribution of various properties, e.g. reliability and relevance, that can be measured via survey data and then used to understand fairness judgements in novel settings. Yaghini et al. [44] proposes an EOP-style model that first identifies circumstance versus effort values using surveys, and then uses a conjoint design to estimate utility. And Lee et al. [22] developed WeBuildAI, an approach built on theory from the collective participation literature, that allows for the measurement and aggregation of scale-based fairness inputs.

Our work emphasizes the need for approaches like these, and extends existing work in several important ways. First and foremost, we cross-compare decisions on multiple services/resources within the same study. In other words, the decision being made is, unlike any prior work, a variable within our experimental design. We find that the decision being made impacts fairness perceptions in both expected and, more notably, unexpected ways. Indeed, there were almost no clear main effects that emerged from our hypothesized 2x2 service setting model. The takeaway is that while others have studied properties of attributes that inform their use in fairness judgements [12], and demographics are salient factors differentiating respondents, *our work shows we have little current understanding of the ways in which we can break down different kinds of services systematically.*

Second, our work finds no systematic effects of priming respondents to consider fairness, relative to simply asking them who "should" get a service. It is of course possible that our fairness manipulation is too weak, or that our sample size is too small. However, if it is indeed the case that "fair" and "should" are largely the same, as has in any case been implicitly assumed in some prior work, then researchers should be cautious about how these judgements are used to inform a machine learning model's decisions. This is because the data on which such models are trained, particularly in settings like Social Work and health, *already* contain decisions on who "should" get services. This is, of course, what a machine learning model seeks to learn. Adding fairness judgements, then, risks simply substituting one definition of "should" for another. At its best, this has the potential to bring power to voices absent in our data and model construction. At worst, it provides yet another lever upon which power inequalities in the machine learning pipeline can be pulled.

Third, while question framing posed limited impacts, we do find at least one obvious difference between the behaviorally-oriented conjoint analysis questions and the more reflexive Likert scale questions we asked. Scholars seeking to use fairness judgements to guide model decisions can learn from this observation. Namely, the goal of soliciting what people "really think" may be best served by a conjoint design. But, using a Likert design may help to enforce social norms that are known to have favorable properties, e.g. in the reduction of prejudice [36]. In the context of studying fairness perceptions, then, it is critical to decide whether we want to understand implicit preferences–in which case one might turn to conjoint approaches– or to solicit normative values– in which case a Likert approach may be more appropriate.

Finally, and related, is that regardless of how one asks questions, fairness perceptions should not be seen as a panacea for creating fair, just, or equitable models. The most obvious example in our work is that among the most consistent effects were 1) a lack of use of race as a decision factor, and 2) a consistently strong use of criminal history. Given the systematic construction of an association between Blackness and criminality [2], a model that uses criminal history as a decision factor is one that further enables white supremacy [16]. Further, Systems Justification Theory suggests that this relationship does not necessarily subside if we solicit fairness perceptions from affected individuals. If we seek to be fair "in an already unjust world" [16], then, fairness perceptions *must* be used in concert with a socio-theoretic understanding of the attributes under study.

## 5 CONCLUSION

The present work conducts a conjoint analysis, with associated Likert-scale questions, to analyze fairness perceptions of a large sample of Mechanical Turk participants across different service allocation contexts and task framings. Put simply, our results provide evidence that the *What* and *Who* of fairness perceptions matter, but that there are limited effects of question framing, at least for high-severity allocation contexts.

Our work does, however, suffer from several limitations. Many of these are similar to those in prior work; in particular, Yaghini et al. [44] list a comprehensive set of limitations including assumptions about linearity of attribute effects, the requirement that respondents are forced to choose between two options they may feel similarly about, and the hypothetical nature of the decisions made. In addition to these limitations noted in prior work, three additional points are of importance here. First, our work is largely exploratory, in that we address a number of different research questions with a single sample. Our goal in doing so was to lay groundwork for future, more directed analysis, but this decision limits the extent to which we explore more mechanistic explanations of respondent behavior. Future work could, for example, better address the latent constructs that differentiate service conditions tested here.

Second, and related, results presented here are necessarily selective - while they show results from our survey data that we believe are most salient and generalizable, we nonetheless leave several important questions unanswered. For example, we do not test whether or not effects of respondent partisanship change for different service allocation contexts. To this end, we provide some additional results in the code and data release for this paper, and provide all data needed to ask these additional questions. Second, our analysis is conducted on a non-representative sample of U.S. adults. While we have tried to focus on results that theories and connections to prior work imply are generalizable, we cannot guarantee this to be true. Thus, more work is needed, for example, to address in a causal fashion why we observe differences between Democrats and Republicans in our sample.

Despite these limitations, however, this work opens a variety of future research questions targeted at the complex interactions between what decisions are being made, who is making them, and how they are being asked to make them. As scholarship in the AIES/FAccT community moves towards the use of subjective perceptions in defining fairness criteria for machine learning models, we believe these questions will be critical to address.

## 6 ACKNOWLEDGEMENTS

The authors would like to thank Atri Rudra, Krithika Raj Dorai Raj, Lalit Jain, Anna Gilbert, Connor Wurst, Yuhao Du, Sauvik Das, and the Anonymous Reviewers for thoughts on prior iterations of this work. All authors were funded by Amazon and the NSF under award NSF IIS-1939579.

## REFERENCES

[1] Lene Aarøe and Michael Bang Petersen. 2014. Crowding Out Culture: Scandinavians and Americans Agree on Social Welfare in the Face of Deservingness Cues. *The Journal of Politics* 76, 3 (July 2014), 684–697. https://doi.org/10/f58t6g
[2] Michelle Alexander. 2020. *The new Jim Crow: Mass incarceration in the age of colorblindness*. The New Press.
[3] Ashtom Anderson, Sharad Goel, Gregory Huber, Neil Malhotra, and Duncan J. Watts. 2014. Political Ideology and Racial Preferences in Online Dating. *Sociological Science* 1 (Feb. 2014), 28–40. https://doi.org/10.15195/v1.a3
[4] Edmond Awad, Sohan Dsouza, Richard Kim, Jonathan Schulz, Joseph Henrich, Azim Shariff, Jean-François Bonnefon, and Iyad Rahwan. 2018. The Moral Machine Experiment. *Nature* 563, 7729 (Nov. 2018), 59. https://doi.org/10.1038/s41586-018-0637-6
[5] Reuben Binns, Max Van Kleek, Michael Veale, Ulrik Lyngs, Jun Zhao, and Nigel Shadbolt. 2018. 'It's Reducing a Human Being to a Percentage': Perceptions of Justice in Algorithmic Decisions. In *Proceedings of the 2018 CHI Conference on Human Factors in Computing Systems*. ACM, 377.
[6] Marilynn B. Brewer. 1999. The Psychology of Prejudice: Ingroup Love and Outgroup Hate? *Journal of social issues* 55, 3 (1999), 429–444.
[7] Michael Buhrmester, Tracy Kwang, and Samuel D. Gosling. 2016. *Amazon's Mechanical Turk: A New Source of Inexpensive, yet High-Quality Data?* American Psychological Association, Washington, DC, US. 139 pages. https://doi.org/10.1037/14805-009
[8] C Bustos Navarrete and F Coutinho Soares. 2020. dominanceanalysis: Dominance Analysis. R package version 1.3. 0.
[9] Rachel Freedman, Jana Schaich Borg, Walter Sinnott-Armstrong, John P. Dickerson, and Vincent Conitzer. 2018. Adapting a Kidney Exchange Algorithm to Align With Human Values. In *Thirty-Second AAAI Conference on Artificial Intelligence*.
[10] Rachel Freedman, Jana Schaich Borg, Walter Sinnott-Armstrong, John P. Dickerson, and Vincent Conitzer. 2020. Adapting a Kidney Exchange Algorithm to Align with Human Values. *Artificial Intelligence* 283 (June 2020), 103261. https://doi.org/10.1016/j.artint.2020.103261
[11] Ben Green and Yiling Chen. 2019. Disparate Interactions: An Algorithm-in-the-Loop Analysis of Fairness in Risk Assessments. In *Proceedings of the Conference on Fairness, Accountability, and Transparency (FAT* '19)*. Association for Computing Machinery, Atlanta, GA, USA, 90–99. https://doi.org/10.1145/3287560.3287563
[12] Nina Grgic-Hlaca, Elissa M. Redmiles, Krishna P. Gummadi, and Adrian Weller. 2018. Human Perceptions of Fairness in Algorithmic Decision Making: A Case Study of Criminal Risk Prediction. In *Proceedings of the 2018 World Wide Web Conference (WWW '18)*. International World Wide Web Conferences Steering Committee, Republic and Canton of Geneva, Switzerland, 903–912. https://doi.org/10.1145/3178876.3186138
[13] Jens Hainmueller, Dominik Hangartner, and Teppei Yamamoto. 2015. Validating Vignette and Conjoint Survey Experiments against Real-World Behavior. *Proceedings of the National Academy of Sciences* 112, 8 (Feb. 2015), 2395–2400. https://doi.org/10.1073/pnas.1416587112
[14] Jens Hainmueller, Daniel J. Hopkins, and Teppei Yamamoto. 2014. Causal Inference in Conjoint Analysis: Understanding Multidimensional Choices via Stated Preference Experiments. *Political analysis* 22, 1 (2014), 1–30.
[15] Galen Harrison, Julia Hanson, Christine Jacinto, Julio Ramirez, and Blase Ur. 2020. An empirical study on the perceived fairness of realistic, imperfect machine learning models. In *Proceedings of the 2020 Conference on Fairness, Accountability, and Transparency*. 392–402.
[16] Jonathan Herington. 2020. Measuring fairness in an unfair World. In *Proceedings of the AAAI/ACM Conference on AI, Ethics, and Society*. 286–292.
[17] Jesse Hoey and Neil J. MacKinnon. 2019. "Conservatives Overfit, Liberals Underfit": The Social-Psychological Control of Affect and Uncertainty. *arXiv:1908.03106 [cs]* (Sept. 2019). arXiv:1908.03106 [cs]
[18] Ben Hutchinson and Margaret Mitchell. 2019. 50 Years of Test (Un)Fairness: Lessons for Machine Learning. *Proceedings of the Conference on Fairness, Accountability, and Transparency - FAT* '19* (2019), 49–58. https://doi.org/10.1145/3287560.3287600 arXiv:1811.10104
[19] John T Jost. 2019. A quarter century of system justification theory: Questions, answers, criticisms, and societal applications. *British Journal of Social Psychology* 58, 2 (2019), 263–314.
[20] John T Jost and Mahzarin R Banaji. 1994. The role of stereotyping in system-justification and the production of false consciousness. *British journal of social psychology* 33, 1 (1994), 1–27.
[21] Christopher Jung, Michael Kearns, Seth Neel, Aaron Roth, Logan Stapleton, and Zhiwei Steven Wu. 2019. Eliciting and Enforcing Subjective Individual Fairness. *arXiv:1905.10660 [cs, stat]* (May 2019). arXiv:1905.10660 [cs, stat]
[22] Min Kyung Lee, Daniel Kusbit, Anson Kahng, Ji Tae Kim, Xinran Yuan, Allissa Chan, Daniel See, Ritesh Noothigattu, Siheon Lee, Alexandros Psomas, and Ariel D. Procaccia. 2019. WeBuildAI: Participatory Framework for Algorithmic Governance. *Proceedings of the ACM on Human-Computer Interaction* 3, CSCW (Nov. 2019), 181:1–181:35. https://doi.org/10.1145/3359283
[23] Thomas J. Leeper, Sara B. Hobolt, and James Tilley. 2020. Measuring Subgroup Preferences in Conjoint Experiments. *Political Analysis* 28, 2 (2020), 207–221.
[24] Weiwen Leung, Zheng Zhang, Daviti Jibuti, Jinhao Zhao, Maximilian Klein, Casey Pierce, Lionel Robert, and Haiyi Zhu. 2020. Race, Gender and Beauty: The Effect of Information Provision on Online Hiring Biases. In *Proceedings of the 2020 CHI Conference on Human Factors in Computing Systems*. 1–11.
[25] Omar Lizardo and Michael Strand. 2010. Skills, Toolkits, Contexts and Institutions: Clarifying the Relationship between Different Approaches to Cognition in Cultural Sociology. *Poetics* 38, 2 (April 2010), 205–228. https://doi.org/10.1016/j.poetic.2009.11.003
[26] Keri Mallari, Kori Inkpen, Paul Johns, Sarah Tan, Divya Ramesh, and Ece Kamar. 2020. Do I Look Like a Criminal? Examining how Race Presentation Impacts Human Judgement of Recidivism. In *Proceedings of the 2020 CHI Conference on Human Factors in Computing Systems*. 1–13.
[27] Ritesh Noothigattu, Snehalkumar S. Gaikwad, Edmond Awad, Sohan Dsouza, Iyad Rahwan, Pradeep Ravikumar, and Ariel D. Procaccia. 2018. A Voting-Based System for Ethical Decision Making. In *Thirty-Second AAAI Conference on Artificial Intelligence*.
[28] Gabriele Paolacci and Jesse Chandler. 2014. Inside the Turk: Understanding Mechanical Turk as a Participant Pool. *Current Directions in Psychological Science* 23, 3 (June 2014), 184–188. https://doi.org/10.1177/0963721414531598
[29] Gabriele Paolacci, Jesse Chandler, and Panagiotis G. Ipeirotis. 2010. Running Experiments on Amazon Mechanical Turk. *Judgment and Decision making* 5, 5

(2010), 411–419.
[30] Michael Bang Petersen. 2012. Social Welfare as Small-Scale Help: Evolutionary Psychology and the Deservingness Heuristic. *American Journal of Political Science* 56, 1 (2012), 1–16.
[31] Michael Bang Petersen, Rune Slothuus, Rune Stubager, and Lise Togeby. 2011. Deservingness versus Values in Public Opinion on Welfare: The Automaticity of the Deservingness Heuristic. *European Journal of Political Research* 50, 1 (2011), 24–52.
[32] Toni Rodon and Marc Sanjaume-Calvet. 2020. How Fair Is It? An Experimental Study of Perceived Fairness of Distributive Policies. *The Journal of Politics* 82, 1 (2020), 384–391.
[33] John E. Roemer and Alain Trannoy. 2016. Equality of Opportunity: Theory and Measurement. *Journal of Economic Literature* 54, 4 (2016), 1288–1332.
[34] Nripsuta Ani Saxena, Karen Huang, Evan DeFilippis, Goran Radanovic, David C. Parkes, and Yang Liu. 2020. How Do Fairness Definitions Fare? Testing Public Attitudes towards Three Algorithmic Definitions of Fairness in Loan Allocations. *Artificial Intelligence* 283 (June 2020), 103238. https://doi.org/10.1016/j.artint.2020.103238
[35] Megha Srivastava, Hoda Heidari, and Andreas Krause. 2019. Mathematical Notions vs. Human Perception of Fairness: A Descriptive Approach to Fairness for Machine Learning. In *Proceedings of the 25th ACM SIGKDD International Conference on Knowledge Discovery & Data Mining (KDD '19)*. Association for Computing Machinery, Anchorage, AK, USA, 2459–2468. https://doi.org/10.1145/3292500.3330664
[36] M.E. Tankard and Elizabeth Levy Paluck. 2016. Norm Perception as a Vehicle for Social Change. *Social Issues and Policy Review* 10, 1 (2016), 181–211.
[37] R Core Team et al. 2013. R: A language and environment for statistical computing. (2013).
[38] Kees Van den Bos. 2003. On the Subjective Quality of Social Justice: The Role of Affect as Information in the Psychology of Justice Judgments. *Journal of personality and social psychology* 85, 3 (2003), 482.
[39] Kees Van den Bos and E. Allan Lind. 2002. Uncertainty Management by Means of Fairness Judgments. (2002).
[40] Wim van Oorschot. 2000. Who Should Get What, and Why? On Deservingness Criteria and the Conditionality of Solidarity among the Public. *Policy & Politics* 28, 1 (2000), 33–48.
[41] Ruotong Wang, F Maxwell Harper, and Haiyi Zhu. 2020. Factors Influencing Perceived Fairness in Algorithmic Decision-Making: Algorithm Outcomes, Development Procedures, and Individual Differences. In *Proceedings of the 2020 CHI Conference on Human Factors in Computing Systems*. 1–14.
[42] Nicholas Winter, Tyler Burleigh, Ryan Kennedy, and Scott Clifford. 2019. *A Simplified Protocol to Screen Out VPS and International Respondents Using Qualtrics*. SSRN Scholarly Paper ID 3327274. Social Science Research Network, Rochester, NY.
[43] Allison Woodruff, Sarah E. Fox, Steven Rousso-Schindler, and Jeffrey Warshaw. 2018. A Qualitative Exploration of Perceptions of Algorithmic Fairness. In *Proceedings of the 2018 CHI Conference on Human Factors in Computing Systems (CHI '18)*. Association for Computing Machinery, Montreal QC, Canada, 1–14. https://doi.org/10.1145/3173574.3174230
[44] Mohammad Yaghini, Hoda Heidari, and Andreas Krause. 2019. A Human-in-the-Loop Framework to Construct Context-Dependent Mathematical Formulations of Fairness. *arXiv:1911.03020 [cs]* (Nov. 2019). arXiv:1911.03020 [cs]